# Precision Measurement of the $D_s^{+*} - D_s^+$ Mass Difference

We have measured the vector-pseudoscalar mass splitting $M(D_s^{*+}) - M(D_s^+) = 144.22 \pm 0.47 \pm 0.37 MeV$, significantly more precise than the previous world average. We minimize the systematic errors by also measuring the vector-pseudoscalar mass difference $M(D^{*0}) - M(D^0)$ using the radiative decay $D^{*0} \to D^0\gamma$, obtaining $[M(D_s^{*+}) - M(D_s^+)] - [M(D^{*0}) - M(D^0)] = 2.09 \pm 0.47 \pm 0.37 MeV$. This is then combined with our previous high-precision measurement of $M(D^{*0}) - M(D^0)$, which used the decay $D^{*0} \to D^0\pi^0$. We also measure the mass difference $M(D_s^+) - M(D^+) = 99.5 \pm 0.6 \pm 0.3$ MeV, using the $\phi\pi^+$ decay modes of the $D_s$ and $D^+$ mesons.

## I. INTRODUCTION

Mass splittings between states with the same quark content but different spin configurations give essential information on the nature of the interquark potential. For example, masses of states with orbital angular momentum can be used to probe the contributions to the potential from spin-orbit and tensor forces between the quarks. High-precision measurements of mass splittings between states without orbital excitation (e.g., the pseudoscalar and vector S-states studied in this paper) give information on the relative contributions of chromoelectric and chromomagnetic terms to the Hamiltonian. Comparing the masses of resonant states having the same spin and charge configuration, but differing in the mass of one of the constituent quarks can isolate the effects of individual terms in the interquark potential. Of particular interest here is the vector-pseudoscalar mass splitting for $c\bar{s}$ mesons compared with $c\bar{d}$ mesons, as these are identical in the flavor SU(3) limit. Differences between them are presumably due to differences in the chromomagnetic contribution to the interquark potential and to the different value of the wave function at the origin, because of the different light quark mass.

Using the decay modes $D^{*+} \to D^+\pi^0$ and $D^{*0} \to D^0\pi^0$, CLEO II recently produced the definitive measurements of the mass splittings between the vector and pseudoscalar non-strange charmed mesons: $M(D^{*+}) - M(D^+)$, and $M(D^{*0}) - M(D^0)$ [?]. These high precision measurements were made possible by: a) the large data sample accumulated by the CLEO II experiment, b) the CLEO II crystal calorimeter, which allowed us to reconstruct the decay mode $D^* \to D\pi^0$ with high efficiency and good resolution, and c) the fact that the decay through pion emission is close to threshold, giving excellent precision on the $D^*$ mass.

Although there were a comparable number of observed events corresponding to the radiative decay $D^{*0} \to D^0\gamma$, this mode was not used because the larger Q-value degrades the mass-difference precision relative to the $D^{*0} \to D^0\pi^0$ mode. However, having measured the $D^{*0} - D^0$ splitting to an accuracy of better than 100 KeV using the pionic mode, we can use this to calibrate the mass difference measurement in the radiative mode. This, in turn, can be used to eliminate many systematic errors in our measurements of $D_s^{*+} \to D_s\gamma$.

## II. DETECTOR, DATA SAMPLE, AND EVENT SELECTION

The CLEO II detector is a general purpose solenoidal magnet spectrometer and calorimeter. Elements of the detector, and performance characteristics, are described in detail elsewhere [?]. The detector is designed to have high efficiency for triggering and reconstruction of both leptonic and hadronic events. Charged particle momentum measurements are made with three nested coaxial drift chambers consisting of 6, 10, and 51 layers, respectively. These chambers fill the volume $r$=3 cm to $r$=1 m, where $r$ is the radial coordinate relative to the beam ($z$) axis. Eleven of the layers in the main 51-layer drift chamber have sense-wires which are slanted relative to the beam axis to give measurements of the coordinate along $z$. More precise measurements of the $z$-coordinate are obtained from cathode pads located at the interfaces of the three tracking chambers. The system achieves a momentum resolution of $(\delta p/p)^2 = (0.0015p)^2 + (0.005)^2$, where $p$ is the momentum, measured in GeV/c. Pulse height measurements in the main 51-layer drift chamber provide dE/dx resolution of 6.5% for Bhabhas, giving good $\pi/K$ separation up to momenta of 700 MeV/c. Outside the central tracking chambers are plastic scintillation counters which are used as a fast element in the trigger system and also give particle identification information from time of flight measurements. The scintillation counters have a resolution of 154 ps as measured for hadrons, allowing better than $3\sigma$ $\pi/K$ separation up to momenta of 1.2 GeV/c.

Beyond the time of flight system is the electromagnetic calorimeter, consisting of 7800 thallium doped CsI crystals [?]. The crystal array gives an energy resolution of approximately 4% at 100 MeV and 1.2% at 5 GeV. The central "barrel" region of the detector covers a solid angle of 75% of $4\pi$. The endcap regions extend the solid angle coverage of the calorimeter to 95% of $4\pi$, although with poorer energy resolution than the barrel region. The tracking system, time of flight counters, and calorimeter are installed in a 1.5 T superconducting coil. Flux return and tracking chambers used for muon detection are located immediately outside the coil and in the two endcap regions.

### A. Data Sample

The data sample consists of $1.70 fb^{-1}$ of $e^+e^-$ annihilations collected at CESR at energies just above and below the $\Upsilon$(4S) resonance, and on the $\Upsilon$(4S) resonance itself; the total data sample corresponds to about $2 \times 10^6$ produced $c\bar{c}$ pairs. All events with 3 or more charged tracks, 1.5 GeV of energy measured in the calorimeter, and having a measured event vertex along the z-coordinate within 5 cm of the known interaction point, are accepted as hadronic event candidates. These events are then used for reconstruction of charmed mesons.

### B. Charged Particle and Neutral Particle Selection

Our study requires the reconstruction of $D_s$ and $D^+$ in the $\phi\pi^+$ mode (with $\phi \to K^+K^-$), and $D^0$ in the $K^-\pi^+$ mode, as well as reconstruction of photons from the radiative transition between the vector and pseudoscalar states. We impose cuts on candidate tracks, requiring mainly that they come from the primary vertex. Candidate charged and neutral particles must satisfy the requirements listed in Table ??. We impose a $\pi^0$ veto on each photon candidate. The $\pi^0$ veto is implemented by matching photon candidates with other photon candidates passing the same quality cuts listed in Table ??. If their invariant mass falls within $2.5\sigma$ (approximately 12 MeV) of the known $\pi^0$ mass and if they give a good kinematic fit to the $\pi^0$ hypothesis, these photons are eliminated from further consideration.

## III. STUDY OF THE $D_S^+$-$D^+$ MASS DIFFERENCE

We begin our study by focusing on the reconstruction of the $\phi\pi^+$ decay mode. We require that the two candidate kaons from the $\phi$ have particle identification information consistent with that expected for real kaons. There is a large background due to uncorrelated $\phi$ and $\pi^+$ candidates, which peaks at $\cos\theta_\phi = 1$, where $\theta_\phi$ is the decay angle of the $\phi$ measured in the charmed-meson rest frame with respect to the charmed-meson momentum vector in the lab frame. The cut $\cos\theta_\phi < 0.8$ is effective in reducing this background while retaining 90% of the isotropic signal.

In decays of pseudoscalar charmed mesons into $\phi\pi^+$, the $\phi$ is polarized and its decay helicity angle (defined as the angle between one of the daughter kaons and the parent charmed-meson in the $\phi$ frame) follows a $cos^2\theta_{helicity}$ distribution. To improve signal-to-noise we require that $|cos\theta_{helicity}| >0.4$. To suppress combinatoric background, we take advantage of the characteristic hard fragmentation function of charmed particles and impose the requirement $x_p(= p_{candidate}/p_{max}) >0.5$. With the above cuts, we obtain the $\phi\pi^+$ invariant mass plot shown in Fig. 1; the mass plot has been fit to two Gaussian signals (representing $D^+ \to \phi\pi^+$, and $D_s \to \phi\pi^+$) on top of a smooth background. The fit to the $D^+ \to \phi\pi^+$ and $D_s \to \phi\pi^+$ peaks yields approximately 400 and 1400 events, respectively.

We use these fitted signals to determine the mass difference between the $D_s$ and $D^+$ mesons. Although uncertainties in the overall mass scale are on the order of 1-2 MeV, we expect the systematic error in the determination of the difference in masses to be much smaller. Contributions to the overall systematic uncertainty may arise from fitting (which we determine to be 0.25 MeV by varying the fit interval and the background function), and from possible differences between the lab momentum spectra of the $\phi$ and $\pi$ daughters in the two cases. We probe the latter effect by fitting the mass difference in bins of scaled momentum $x_p$, as shown in Fig. 2. The data are consistent with no variation as a function of $x_p$ at the 67% confidence level, and we attribute a systematic error less than 0.1 MeV due to such a dependence. We arrive at a total systematic error of 0.3 MeV, and are therefore able to determine the difference in masses

3030194-001

FIG. 1. Invariant mass of $\phi\pi^+$ combinations. The smaller peak is the $D^+ \to \phi\pi^+$ signal and the larger peak is the $D_s^+ \to \phi\pi^+$ signal.

3030194-002

FIG. 2. Mass difference between $D^+$ and $D_s^+$, where both mesons are observed in the $\phi\pi^+$ mode, as a function of scaled momentum $x_p$.

between the $D_s$ and $D^+$ to be 99.5±0.6±0.3 MeV. This value compares well with the present Particle Data Group value of 99.5±0.6 MeV [?].

## IV. DETERMINATION OF THE $D_S^{*+}$-$D_S$ MASS DIFFERENCE

With our large sample of $D_s^+ \to \phi\pi^+$, we can make a precise measurement of the $D_s^{*+}$-$D_s$ mass difference. The previous best measurement of this mass difference was made by the ARGUS collaboration who obtained a value of 142.5±0.8±1.5 MeV [?]. The statistical precision of their measurement was limited by their small sample of $D_s^{*+} \to D_s\gamma$ events in which the photons converted to $e^+e^-$ pairs. Their systematic precision was limited by the relatively large uncertainty in the calibration mode $D^{*0} \to D^0\gamma$.

The CLEO measurement is made in the following manner. First, we combine $D_s^+$ candidates with photon candidates and use the resulting $D_s^{*+} \to D_s\gamma$ signal to measure the mass-difference $\Delta_s^\gamma$=M($D_s^{*+}$)-M($D_s^+$), where the $s$ subscript on $\Delta$ indicates that we are considering the $c\bar{s}$ meson, and the $\gamma$ superscript indicates that the measurement is made using photon transitions. This raw mass-difference is still susceptible to errors in the overall photon energy calibration[1] which may be effectively eliminated as follows. Using the photon transition $D^{*0} \to D^0\gamma$, we similarly measure $\Delta_u^\gamma = M(D^{*0}) - M(D^0)$, which allows us to calculate the difference between the two mass-differences $\delta$M=$\Delta_s^\gamma - \Delta_u^\gamma$. This may then be used with the high precision measurement of $\Delta_u^\pi$ (using $D^{*0} \to D^0\pi^0$) [?] to obtain $\Delta_s = \Delta_u^\pi + \delta M$.

By imposing the same photon requirements in our measurements of the two radiative transitions under consideration we can extract $\Delta_s$ relatively free of uncertainties in the absolute photon energy calibration. This technique is limited largely by differences in fitting the two signals due to the presence of the large $D^{*0} \to D^0\pi^0$ feed-down in the $D^0\gamma$ mass-difference plot. There are no hadronic decays of the $D_s^{*+}$ states to produce such a reflection in the $D_s^{*+} \to D_s\gamma$ mass-difference plot.

### A. Measurement of $\Delta_s^\gamma$

As discussed above, we reconstruct $D_s^{*+}$'s in the mode $D_s^{*+} \to D_s\gamma$. In order to improve the signal-to-noise we cut on the decay angle $\theta_\gamma$ of the photon in the $D_s^{*+}$ frame. Requiring cos$\theta_\gamma$ > -0.7 eliminates a significant background, as is evident from Fig. 3. This requirement is made in addition to the other photon cuts detailed in Table ??.

Figure 4 shows the distribution we obtain for $M(\phi\pi^+\gamma) - M(\phi\pi^+)$. The mass-difference distribution is fit to the sum of a smooth polynomial plus a "Crystal Ball Line Shape"[2] around the region of the expected signal.[3] The width of the signal and the magnitude of the tail are set at values obtained from Monte Carlo simulations. The mass difference we obtain from this direct measurement is 144.70±0.42 MeV, where the error is statistical only.

### B. Measurement of $\Delta_u^\gamma$ and Determination of $\delta M$

We reconstruct $D^0$'s in the mode $D^0 \to K^-\pi^+$. The mass-difference signal $M(D^0\gamma) - M(D^0)$ is shown in Fig. 5 with two different fits overlaid. To obtain this mass-difference plot, we used the same photon cuts as in the $D_s^{*+} \to D_s\gamma$ analysis. As before, we perform a fit (Fig. 5a) using the Crystal Ball Line shape function plus a smooth background. We explicitly exclude the low mass enhancement from $D^{*0} \to D^0\pi^0$ from the fit region. The mass difference obtained is $\Delta_u^\gamma = 142.61 \pm 0.21$ MeV (statistical errors only). This

[1]The photon energy calibration is based on fitting the observed $\pi^0$ mass peak over a wide range of $\pi^0$ momenta.

[2]The Crystal Ball Line Shape is a nearly Gaussian distribution with a tail on the low end to take into account processes which may give an undermeasurement of the true photon energy.

[3]The enhancement at low mass-difference arises from misidentified $D^*+ \to D^+\pi^0$ events where the $D^+$ decays to a three-body final state such as $K^-\pi^+\pi^+$. When one of the final state particles is misidentified, kinematic reflections can occur in a mass region around the $D_s \to \phi\pi^+$ signal. This has been verified by examining mass-differences using $\phi\pi^+$ combinations from the $D_s$ sideband region.

3030194-010

FIG. 3. M($\phi\pi^+\gamma$)-M($\phi\pi^+$) vs. cos$\theta_\gamma$ where $\theta_\gamma$ is the photon emission angle in the $\phi\pi^+\gamma$ frame relative to the $\phi\pi^+\gamma$ direction in the lab. Transition photon candidates are required to have cos$\theta_\gamma$ > -0.7, as described in the text.

3030194-007

FIG. 4. Mass difference between $D_s\gamma$ and $D_s$ with fit overlaid.

may be compared with the value obtained from the $\pi^0$ transition, $\Delta_u^\pi = 142.12 \pm 0.05 \pm 0.05$ MeV, indicating that our overall photon energy calibration is understood to within 0.5% for the photons of interest in this measurement. To test fitting systematics, we perform an additional fit to this mass-difference plot (shown in Fig. 5b), where we explicitly account for the reflection from the hadronic mode. This is detailed further in our discussion of systematic errors.

Comparing the mass differences obtained from Fig. 5a and Fig. 4, we determine $\delta$M=$\Delta_s^\gamma - \Delta_u^\gamma$ = 2.09$\pm$0.47 MeV as summarized in Table ??. Combining this value with $\Delta_u^\pi$ gives $\Delta_s$ = 144.22$\pm$0.47 MeV. The errors quoted in both numbers are statistical errors only.

### C. Systematic Errors

Systematic uncertainties arise from sources which affect $\Delta_s^\gamma$ and $\Delta_u^\gamma$ differently and therefore introduce shifts in $\delta M$. To the extent that the $D_s^{*+}$ and $D^{*0}$ fragmentation functions are different, photon energy calibration uncertainties can introduce systematic shifts, although the good agreement between $\Delta_u^\gamma$ and $\Delta_u^\pi$ indicates that the photon energy scale is relatively well-understood. As is evident from Figs. 4 and 5, the background shapes are different in the two cases and there are therefore additional uncertainties arising from signal extraction systematics.

We have studied possible biases using Monte Carlo simulations. Given input values of $M(D^{*0}) - M(D^0)$ and $M(D_s^{*+}) - M(D_s^+)$ we are able to recover values which are consistent with the input numbers after processing the Monte Carlo data through our analysis software. For the $D_s^{*+} \to D_s\gamma$ transition, for example, inputting a mass difference between $D_s^{*+}$ and $D_s$ of 142.60 MeV, we recover a value of 142.55$\pm$0.15 MeV.

We have investigated the dependence of the measured mass-difference on the photon energy and on the momentum of the $D_s^{*+}$, which is correlated with the photon energy. Figure 6 demonstrates that the dependence of the measured mass-difference on transition photon energy is not large. Figure ?? shows the measured mass difference as a function of the scaled momentum $x_p$ of the $D_s^{*+}$. The plot is consistent with no variation of mass difference with momentum. We therefore attribute no additional systematic error to such sources.

There is also an uncertainty of $\pm$0.5% in the absolute photon energy calibration which results in an error of $\pm$0.7 MeV in $\Delta_s^\gamma$ and $\Delta_u^\gamma$ as shown in Table II. However the contribution to $\delta$M=$\Delta_s^\gamma - \Delta_u^\gamma$ is only $\pm$0.02 MeV since the systematic errors essentially cancel each other.

Although systematics due to uncertainties in the overall energy calibration largely cancel, fitting systematics remain. For the signal parameterization, we have checked that variations of signal shape produce shifts in both $\Delta_s^\gamma$ and $\Delta_u^\gamma$ which track each other and therefore cancel in the value of $\delta M$. The presence of the low mass enhancement due to the hadronic decay $D^{*0} \to D^0\pi^0$ can distort the shape of the background[4] in the case of the calibration mode $D^{*0} \to D^0\gamma$. We have done a variety of fits using different assumptions for the photon line shape plus the possible background shapes in order to quantify the extent to which the hadronic decay can change the value of the mass difference we derive. Such a distribution is shown as the overlaid histogram in Fig. 5b. In this case, we have fit our observed signal to a sum of three pieces: a) a mass-difference background (whose shape is obtained from Monte Carlo studies) due to feed-down from $D^{*0} \to D^0\pi^0$, $\pi^0 \to \gamma\gamma$, where one of the $\pi^0$ daughter photons is reconstructed and the second is not detected in the calorimeter, b) a signal representing $D^{*0} \to D^0\gamma$, whose shape was also determined by Monte Carlo simulation, and c) a mass-difference background, obtained from $M(K^-\pi^+\gamma) - M(K^-\pi^+)$, where the $K^-\pi^+$ combination is taken from the $D^0$ sideband regions. This gives a good fit to the data, indicating that we are able to account for the various components of the observed mass-difference plot. From this fit (Fig. 5b), we obtain a value of the mass difference $\Delta_u^\gamma$ of 142.75$\pm$0.24 MeV. This compares well with the mass difference of 142.61$\pm$0.21 MeV obtained from Fig. 5a. We assign a systematic error contribution of 0.3 MeV to the measurement of $\Delta_u^\gamma$ and 0.2 MeV to $\Delta_s^\gamma$, and conservatively assume the errors are totally uncorrelated in determining the contribution to the overall systematic error in $\delta M$.

The results of these measurements are summarized in Table ??.

---

[4]Note, however, that the hadronic mode is kinematically prohibited from producing background in the region of the $D^{*0} \to D^0\gamma$ mass-difference signal.

3030194-003

3030194-006

FIG. 5. $D^{*0} - D^0$ mass difference distribution. a) Shows a fit to the signal expected from true $D^{*0} \to D^0\gamma$ plus a smooth background, as done with $D_s^{*+} \to D_s\gamma$, and b) shows a fit to contributions arising from true $D^{*0} \to D^0\gamma$, $D^{*0} \to D^0\pi^0$, and random photon plus fake $D^0$ combinations, as described in text.

3030194-004

FIG. 6. $D^{*0} - D^0$ mass difference as a function of photon energy.

## V. DETERMINATION OF UPPER LIMIT ON $D_S^{*+}$ WIDTH

It is straightforward to determine a limit on the intrinsic width of the $D_s^{*+}$ meson. The measured upper limit on the intrinsic width of the $D^{*0}$ is $\Gamma < 2.1$ MeV [?]. If we perform a free fit to the mass-difference signals observed in $D^{*0} \to D^0\gamma$ (Fig. 5) and $D_s^{*+} \to D_s\gamma$ (Fig. 4), using the same signal shape but allowing the width of the photon peak to vary, we obtain values of 4.50±0.24 and 4.29±0.40 MeV for the widths of the respective signals. Relating this to the intrinsic and experimental widths of the two resonances, we have:

$$\sqrt{\sigma^2_{D^{*0}} + \sigma^2_{exptl}} = 4.50 \pm 0.24 MeV, \sqrt{\sigma^2_{D_s^*} + \sigma^2_{exptl}} = 4.29 \pm 0.40 MeV. \quad (1)$$

Assuming that the experimental resolutions $\sigma_{exptl}$ are identical for $D^{*0} \to D^0\gamma$ and $D_s^{*+} \to D_s\gamma$, we can square and subtract these two expressions to obtain $\Gamma_{D_s^*} < 4.9$ MeV at 90% confidence level. This technique is, at present, limited by the statistical precision on the $\Delta^\gamma$ measurements.

## VI. SUMMARY

We have made a new measurement of the mass difference between the $D_s$ and the $D^+$ mesons, obtaining a value (99.5±0.6±0.3 MeV) in good agreement with the present world average, and with comparable errors. Calibrating our $D_s^{*+}$-$D_s$ mass-difference using the mass-difference observed in the $D^{*0} \to D^0\gamma$ mode, we determine $\delta$M=$\Delta_s^\gamma - \Delta_u^\gamma$=2.09±0.47±0.37 MeV. Combining this value with our previous measurement of the $D^{*0} - D^0$ mass difference [?], we determine M($D_s^{*+}$)-M($D_s^+$)=144.22±0.47±0.37 MeV. This value is much more precise than the previous world average of 142.4±1.7 MeV [?].

It is of interest to compare the vector-pseudoscalar mass splitting for the $c\bar{s}$ system with that of the $c\bar{d}$ system. Two factors in the expression for the mass difference depend on the mass of the light quark: (i) the chromomagnetic effect is expected to be smaller for the $c\bar{s}$ system due to the heavier strange quark, but (ii) the square of the wave function overlap at the origin is expected to be larger because of the larger reduced mass of the strange quark. Our measurements indicate a larger vector-pseudoscalar splitting in the $c\bar{s}$ system than in the $c\bar{d}$ system, indicating that wave function overlap is the dominant effect.

Finally, using the signal we observe in both the $D^{*0} \to D^0\gamma$ and $D_s^{*+} \to D_s\gamma$ modes, we determine the intrinsic full width of the $D_s^{*+}$ to be <4.91 MeV at 90% confidence level.

Table ?? summarizes the vector and pseudoscalar splittings obtained by this and previous measurements.

## ACKNOWLEDGEMENTS

We gratefully acknowledge the effort of the CESR staff in providing us with excellent luminosity and running conditions. This work was supported by the National Science Foundation and the U.S. Dept. of Energy.

---

3030194-005

FIG. 7. $D^{*0} - D^0$ mass difference as a function of scaled $D^0$ momentum ($x_p$).

| | |
|---|---|
| DOCA for all tracks in $\|r-\phi\|$ | <5 mm |
| DOCA for all tracks in $\|r-z\|$ | <5 cm |
| \|dE/dx\| deposition for tracks | $< 2\sigma/3\sigma$ from expected for K/$\pi$ |
| Measured $\phi(\rightarrow K^+K^-)$ mass | $\pm 2.5\sigma$ (10 MeV) of known mass |
| Measured $D_s^+(\rightarrow \phi\pi^+)$ mass | $\pm 2.5\sigma$ (21 MeV) of known mass |
| Measured $D^0(\rightarrow K^-\pi^+)$ mass | $\pm 2.5\sigma$ (26 MeV) of known mass |
| Charmed meson momentum | $x_p > 0.5$ |
| $\|cos\theta\|$ for photon candidates | <0.7 (barrel region) |
| Photon candidates | unmatched to charged tracks |
| Photon shower isolation | >50 mrad from other showers |
| Photon energy | > 50 MeV |
| Photon lateral energy deposition | 99% probability of coming from true photons |
| Photon $\pi^0$ veto | inconsistent with coming from $\pi^0 \rightarrow \gamma\gamma$ |

TABLE I. **Summary of Cuts used in the Analysis. ("DOCA" denotes distance of closest approach to the interaction point.)**

| | $\Delta_s^\gamma$ | $\Delta_u^\gamma$ | $\delta$M=$\Delta_s^\gamma - \Delta_u^\gamma$ |
|---|---|---|---|
| Raw $\Delta$M | 144.70 | 142.61 | 2.09 |
| Statistical Error | ±0.42 | ±0.21 | ±0.47 |
| Signal width systematic | 0.1 | 0.03 | 0.1 |
| Signal tail systematic | 0.09 | 0.06 | <0.05 |
| Momentum cut systematic | - | - | - |
| $cos\theta_\gamma$ cut systematic | - | - | - |
| Background fit systematic | 0.3 | 0.2 | 0.36 |
| Absolute $E_\gamma$ calibration systematic | ±0.7 | ±0.7 | ±0.02 |

TABLE II. **Summary of Mass-Difference Results. (All numbers are in MeV.)**

| | |
|---|---|
| M($D^+ - D^0$) | 4.79±0.10 MeV [?,?] |
| M($D_s^+$-$D^+$) | 99.5±0.67 MeV [?],[this measurement] |
| M($D^{*+} - D^{*0}$) | 3.32±0.08±0.05 MeV [?] |
| M($D_s^{*+}$-$D_s^+$)-M($D^{*0} - D^0$) | 2.09±0.47±0.37 MeV [this measurement] |
| M($D_s^{*+}$-$D_s^+$) | 144.22±0.47±0.37 MeV [this measurement] |

TABLE III. **Summary of Charmed Meson Mass Splittings**